# Towards a Novel Cooperative Logistics Information System Framework

Fares Zaidi [1], Laurent Amanton [2], Eric Sanlaville [3]

[1,2,3] Normandie Univ, UNIHAVRE, LITIS, 76600 Le Havre, France
{fares.zaidi, laurent.amanton, eric.sanlaville}@univ-lehavre.fr

**Abstract.** Supply Chains and Logistics have a growing importance in global economy. Supply Chain Information Systems over the world are heterogeneous and each one can both produce and receive massive amounts of structured and unstructured data in real-time, which are usually generated by information systems, connected objects or manually by humans. This heterogeneity is due to Logistics Information Systems components and processes that are developed by different modelling methods and running on many platforms; hence, decision making process is difficult in such multi-actor environment. In this paper we identify some current challenges and integration issues between separately designed Logistics Information Systems (LIS), and we propose a Distributed Cooperative Logistics Platform (DCLP) framework based on NoSQL, which facilitates real-time cooperation between stakeholders and improves decision making process in a multi-actor environment. We included also a case study of Hospital Supply Chain (HSC), and a brief discussion on perspectives and future scope of work.

## 1. Introduction

Supply chains are the backbone of the global economy. Nowadays, research and innovation have become a major asset for improving sustainability and performance of supply chains. Needing for design, manage, and operate these complex global supply chains continues to grow. Basically, Supply Chain Management (SCM) is the art of delivering products from manufacturers to customers. This definition can also be extended to service industries like tourism and health care which have some of the most specialized Supply Chains.
Managing supply chains involves dealing with four flows: the flow of physical products, the flow of money, the flow of information, and the reverse flow of products at the end of their life for recycling, re-manufacturing or disposal. In our study, we focus on the flow of information or data.

Logistics corridor such as Seine Valley gathers many socio-economic stakeholders, each one with his information system (IS). According to Chen et al. [1], performance factors that influence Supply Chains (SCs), in particular hospital SCs are: data confidentiality, information exchange and integration between supply chain actors' distributed and heterogeneous Information Systems. Thus, improving performance of a distributed SC implies the design of a system that guarantees and satisfies these performance indices, while taking into account the evolution of exchanged data flows, and the increasing number of stakeholders (scalability).

Logistics information are heterogeneous and fragmented [2] [3] since each stakeholder has his own Logistics Information System, usually an Enterprise Resource Planning (ERP). Thus, several challenges have arisen. Firstly, the most part of generated and exchanged data is unstructured and therefore difficult to store, process or analyze with legacy tools. Secondly, the competition between logistics actors or companies involves a lack of information sharing, cooperation and reactivity to risks. Although these actors have often divergent interests, they also need each other's support. Hence, it is more than ever necessary to define a new paradigm for Logistics Information Systems (LIS). This paradigm consists of enhancing flexibility and scalability of the LIS [4], handling massive and heterogeneous data, enabling a better real-time cooperation between stakeholders and facing reactively to risks while taking into account data and confidentiality issues. The main goal is to improve decision making process of all supply chain actors.

In this paper, we propose a novel Distributed Cooperative Logistics Platform (DCLP) framework, which relies on the emerging NoSQL databases that appears as the appropriate tools to meet our needs. The rest of this paper is organized as follows. Section 2 presents Logistic Information Systems and NoSQL Databases. Then, context and our NoSQL based LIS architecture are described in section 3, as well as a case study of Hospital Supply Chain. Finally, we conclude the paper in section 4, and highlight future scope of work.

## 2. Information Systems for cooperative Supply Chains

### 2.1. Supply Chains and information management tools

Supply chain as defined by Cristopher [5] is "*a network of organizations that are involved through upstream and downstream linkages in the different processes and activities that produce value in the form of products and services in the hands of the ultimate customer*".

In fact, Supply Chains are composed of multiple stakeholders like suppliers, manufacturers, carriers, distributors and customers. As shown in Figure 1, Supply Chain deals with four flows, which are physical, money, data and reverse flows. Furthermore, all stakeholders or organizations have Plan [6], source and deliver activities, and most of them perform make activity, except some ones (like carriers or distributors).

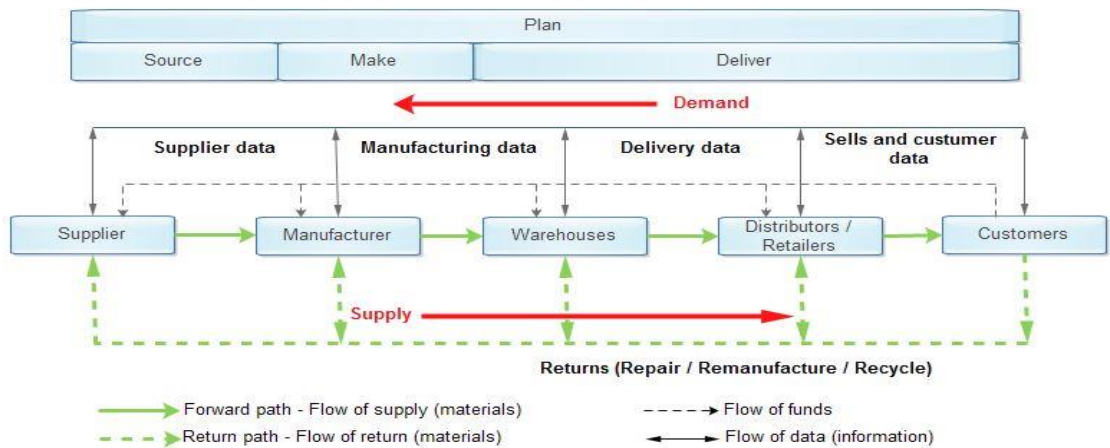

**Figure 1:** General Supply Chain (adapted from [6]).

Organizations and their activities within the supply chain are integrated thanks to Plan activity [6]. If we look at stakeholders of a supply chain with a macro process view, we can see that each member of the Supply Chain has three main components or modules, as illustrated in Figure 2 below.

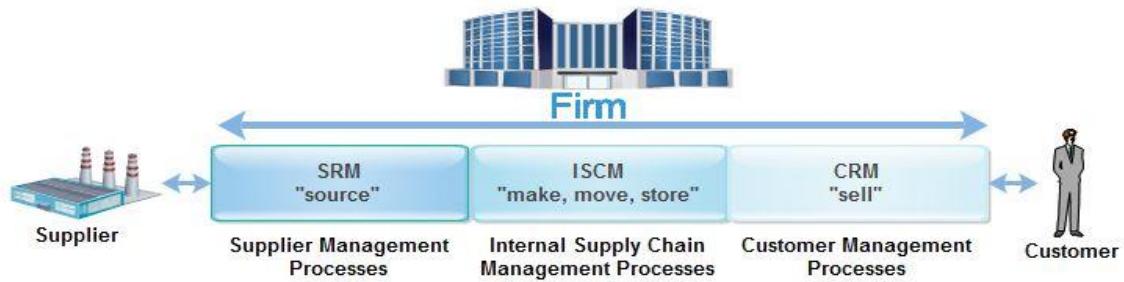

**Figure 2:** Macro perspective of a company in a Supply Chain (adapted from [7]).

Companies connects to their suppliers with source activities (like acquiring raw materials) using Supplier Resource Management (SRM) module, and to their customers with deliver activities (like delivery of goods) using Customer Resource Management (CRM) module. They also operate their internal actions or processes like make, move and store using Internal Supply Chain Management processes (ISCM) [8]. These different functions need to communicate with each other in order to exchange data easily, securely, correctly and in real-time in some cases.

Supply Chains are very heterogeneous and complex, they involve multiple players which need instantaneous communications for B2B (Business to Business), B2C (Business to Customer) and M2M (Machine to Machine). Logistic Information Systems represent the data side of Supply Chains; their main objective is to manage information flows through Supply Chains.

Managing data and information is essential to control other flows, i.e. physical, money and reverse flows. Hence, Logistics Information Systems have an important role in Supply Chain Management since they are the hub of interconnection between different flows and their related platforms. There are three classes of ICT systems related to logistics:

- Enterprise Resource Planning (ERP): represents the main database for firm activity.
- Supply Chain Planning: consists of Product Lifecycle Management (PLM), scheduling and production Planning.
- Supply Chain Execution: Warehouse Management Systems (WMS), Manufacturing Execution Systems (MES) and Transportation Management Systems (TMS).

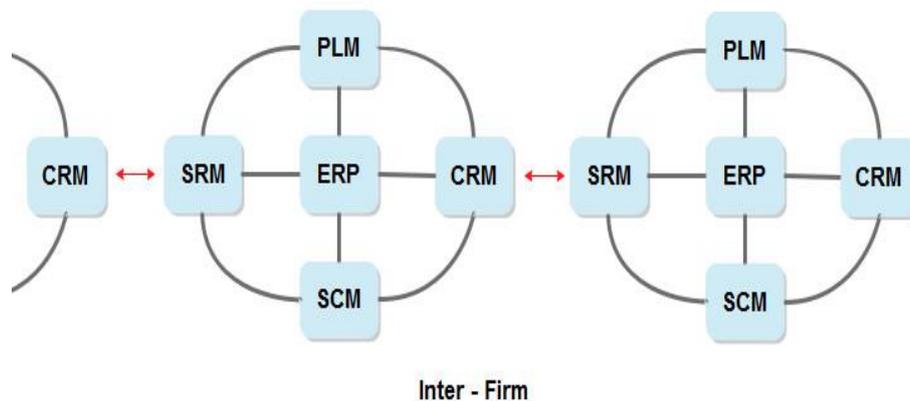

**Figure 3**: General Inter-firms interaction.

Figure 3 represents interconnection between Logistics companies. We can see that The Customer Relationship Management module of each company interacts with the Supplier Relationship management module of the other. This interconnection may result in integration issues among separately designed systems, in particular real-time managing of information flows. H. Wortmann and al. [9] organized interviews with 9 distribution companies specialised in Cross-Docking and they noted the following problems:

- GPS systems used to track merchandise and packages can't provide information about causes of delays and other issues. Hence, planners tend to communicate with drivers by email or mobile in order to be informed about transportation state and eventual issues and solutions.
- In addition to the previous problem, GPS systems are not integrated with stakeholders' transactional systems (like ERP or WMS), which implies that the last ones are not always up to date in real-time.
- Decision Support Systems (DSS) are not reliable as they are fed by transactional systems which can store outdated data.
- The multi-stakeholder environment accentuates previous issues.

We can see here that there is a vicious circle created by issues related to GPS, transactional and decision support systems. Hence, it is necessary to break this circle and improve real-time cooperation and integration between heterogeneous LISs.

A Virtual e-Chain model (VeC) for Supply Chain collaboration in real-time was proposed by V. Manthou and al. [10], it aims to improve integration and collaboration between Supply Chain partners in order to anticipate dynamic customers requirements and enable intelligent decisions based on knowledge acquisition. VeC framework is divided into four modules:

- e-Supply chain integration module which facilitates data and transaction flow between actor's software applications.
- e-Supply chain process modelling module that uses complex e-business models and rules to develop roles specifications, like activities and responsibilities.
- e-Supply chain partners relationship management (e-PRM) module that enables the collaborative management and monitoring of supply chain stakeholders.
- e-Supply chain intelligence module that tracks collaborative channel events and presents decision-oriented information in order to improve decision making process.

Once VeC partners enter into a business relationship, mutual success will depend on mutual trust. In reality, most of Supply Chains partners are reluctant to share their information with each other directly, especially with their direct competitors, this is why it is difficult to implement this model. The framework that we propose in section 3 can handle this privacy issue.

Usually, Frequent information exchange is required between Supply Chain stakeholders to streamline traffic and improve decision-making process. For example, a hospital can organize itself more effectively internally (scheduling operations, scheduling appointments, receiving as many patients as possible and ensure their comfort) by receiving as much external information as possible, like state and delivery time of medications or consumables, with a possible confidence interval, or like searching for supplier that has available stock and who can deliver the desired product before others; the main way to get such information is to email or call suppliers, carriers, and to verify road traffic conditions. We can notice that this task is time consuming, especially since it must be done daily, this can cause uncertainties and unforeseen delivery delays. Another example is supplier who want to predict future orders, he need to get some information from his customers to perform this task in order to improve production process.

In addition to the fact that data exchanged between Supply Chain stakeholders is heterogeneous and unstructured, it becomes difficult to deal with different risks and challenges mentioned above without help of an automated collaborative system. However, the major difficulty that appears while collaborating in an inter-enterprise environment is the guarantee of data confidentiality and reluctance of each actor to share information with others, especially if it concerns its competitors, this can be a serious obstacle to setting up such cooperative platform [11]. Furthermore, real-time information flows exchanged between actors are characterized by two important aspects: data freshness (delivery, geolocation, etc.), and transactions deadline.

The inter-firm collaborative platform that we propose (DCLP) can handle all these constraints, i.e. storing and handling massive amount of heterogeneous shared data in real-time, and guaranteeing confidentiality of actor's data. After research and literature review, we found that NoSQL databases are the best suited IT technology to satisfy these constraints. We give an overview of these NoSQL systems in section 2.2.

**2.2. NoSQL Databases**

Non-relational, open source and distributed databases also known as Not Only SQL or NoSQL are new generation databases that can handle massive amounts of data in real-time, whether structured, semi-structured or unstructured data [12]. They appeared by 2007 when Google and Amazon engineers designed BigTable [13] and DynamoDB [14] respectively.
NoSQL databases are schema-less, i.e. their schema is dynamic and flexible, this is very useful to store and handle a variety of data structures, unlike legacy SQL or Relational Database Management Systems (RDBMS) that have a rigid schema; Moreover, these latter are based on ACID properties (Atomicity,

Consistency, Isolation, Durability). Firstly, Atomicity requires that each transaction must be indivisible or atomic, if one slice of a transaction fails, then the whole transaction fails. Secondly Consistency ensures that the data stored in a cluster must be identical into all nodes, when a write is performed to one node, this write must be copied to all other nodes before responding to a request. Thirdly, Isolation stipulates that each transaction has to be executed as if it was the only one in the system, and the result of concurrent execution of transactions should be the same as if they were executed sequentially. Finally, Durability ensures that once a transaction is committed, it will be stored in a non-volatile memory before responding to a request, in order to avoid data-loss. All these constraints complicate the introduction of additional nodes (horizontal scalability) and therefore make traditional RDBMS clusters hard to scale. Table 1 summarizes comparison between SQL and NoSQL databases.

Table 1: SQL vs NoSQL databases.

| SQL | NoSQL |
| --- | --- |
| Relational | Non-Relational |
| SQL language | No common language |
| ACID transactions | BASE |
| Hard scalability | Horizontal scalability |
| Not adapted to big data | Performance/ Big data |
| Rigid schema | Schema-less/ Flexible |
| Structured data | Structured and unstructured data |

Thus, to increase write operations capacity and improve cluster scalability, NoSQL databases relax ACID requirements by using CAP theorem (Consistency, Availability, Partition tolerance) and BASE model (Basically Available, Soft-State, and Eventual Consistency).

CAP theorem has been proposed by Eric Brewer in 2000 [15] and formally proven by Seth Gilbert and Nancy Lynch in 2002 [16]. It states that at a specific given moment $t$, a distributed system can only guarantee two of these three constraints:

- Data Consistency (C): All system nodes have always the same view of the data
- Availability (A): guarantee that each client can always read and write
- Partition tolerance (P): the system works well despite physical network partitions

BASE model [17] improves scalability by relaxing consistency constraints (Eventual Consistency) [18] [19], this model is optimistic. Conversely, ACID Model is pessimistic, i.e. it guarantees consistency whatever the circumstances.

NoSQL Databases are classified into four categories or families [20]:

- Key/value databases: key/value pairs can be stored in tables. this data structure reduces data access complexity to O(1)
- Column databases: data are stored in columns, unlike classic RDBMSs which stock data in rows
- Document databases: they use documents called JSON (JavaScript Object Notation) to store data. These documents can be nested or grouped together into a set called Collection
- Graph databases: information are stored into graphs, more exactly data are stored into nodes, and relationship between these nodes appears on edges. Thus, it is possible to query the database using common algorithms of graph theory. Low level layer of these databases can be key/value or document type

Each database from the previous families can be used to operate with specific applications accordingly to several real constraints and needs, for example graph based databases are suitable to be used to store relationship between users in social networks. In a multi-actor logistic environment, data is generally unstructured and distributed over the supply chain, and the system needs to be scalable. So the ideal is to use a NoSQL database. After literature review [20] [21] [22], we decided to use the most popular and powerful Nosql data store, MongoDB.

MongoDB is a document based NoSQL database, it respects the CP (Consistency and Partition tolerance) properties of the CAP theorem. MongoDB can scale horizontally with automated partitioning, also called Sharding, where each shard represents a replica set, and each replica set is composed of one Master (primary) and several slaves (secondaries). Master and slaves are considered as Mongo Daemon instances (mongod); thus, all writes go to the Master by default, so, the master-slave replication is used to ensure a high availability [12]. Figure 4 represents the general framework of MongoDB.

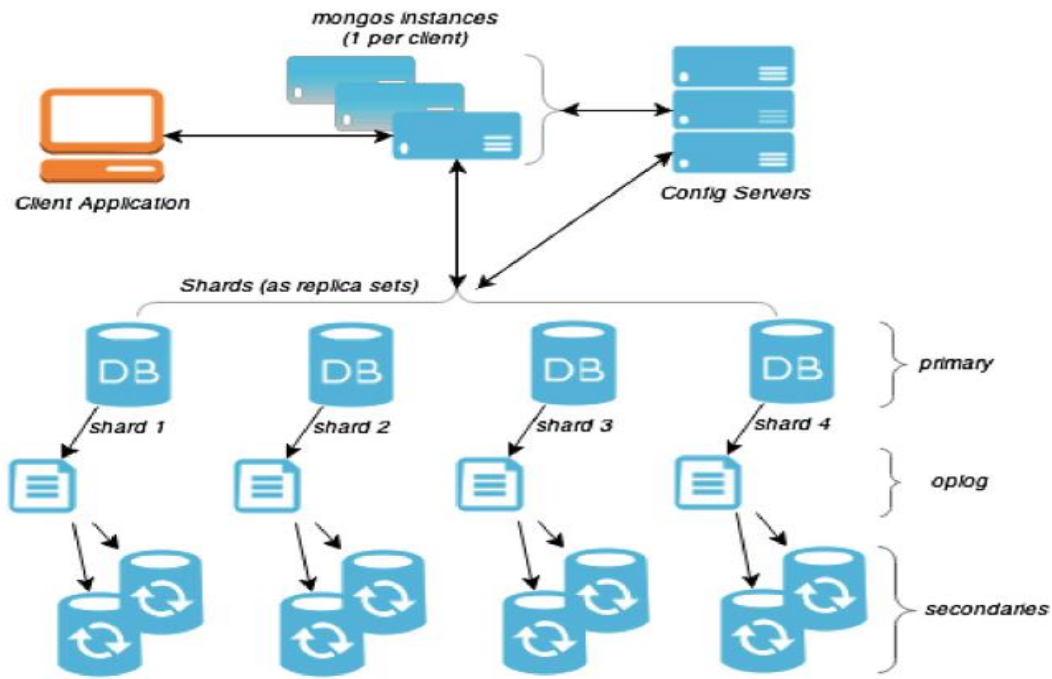

**Figure 4:** MongoDB framework [23].

MongoDB Shard (MongoS) represents the client side of MongoDB, it is required to distribute a data of a sharded collection across a sharded clusters, this process is called Balancing. MongoS acts also as a router in order to link between the customer's application layer and the sharded cluster in order to determine the location of the data.

Oplog (Operations LOG) is a special collection that keeps record of all operations received by the Master, then slaves copy and apply these operations in an asynchronous process in order to maintain the current state of the database. Oplog's size is 5% of physical memory in Linux and windows systems.

The next section describes how we used MongoDB to empower our Logistic Information System framework.

## 3. Approach

### 3.1. Proposed Distributed Cooperative Logistics Platform (DCLP)

In order to face challenges mentioned in section 2.1, we proposed the following solution:
First of all, we proposed to implement an overlay on top of actors' logistics Information Systems, which can store and manage large amount of distributed and heterogeneous data, whether structured or unstructured. NoSQL DBs satisfy well this criterion.
Then, we proposed to use a frontend server hosted by a trusted third party, to aggregate data and reply to Supply Chain actors' queries, and to guaranty data confidentiality.
Finally, from the two previous steps we proposed an architecture of a Distributed Cooperative Logistics Platform (DCLP), that allows actors to share data necessary for functioning of this DCLP.

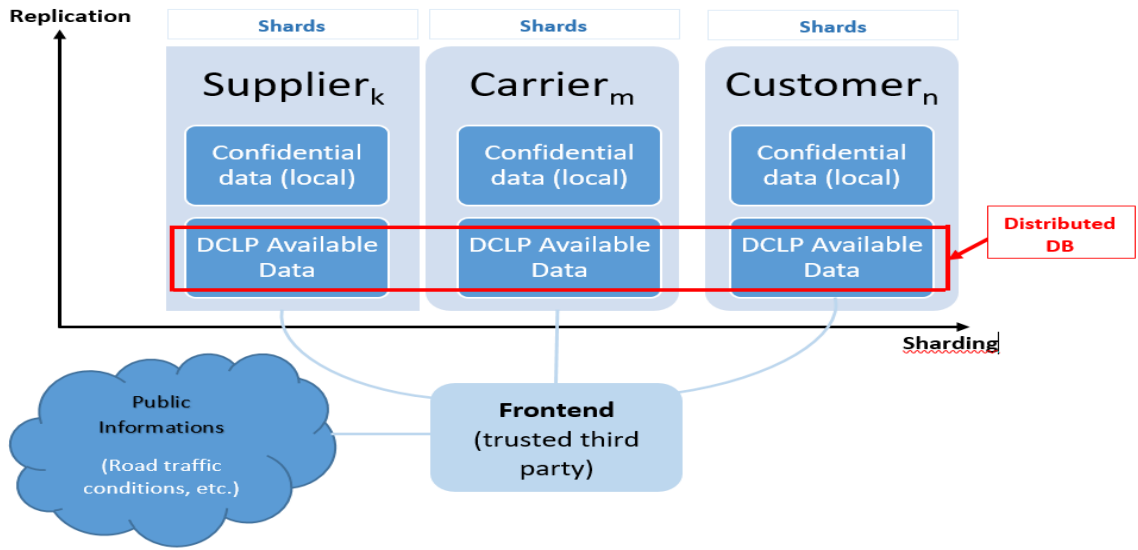

**Figure 5:** Proposed Distributed Cooperative Logistics Platform (DCLP)

Figure 5 illustrates the DCLP, in which each actor has his own Information System, that includes confidential data stored locally, and available data stored into DCLP's shards (parts of the distributed DB). The set of Shards constitutes a distributed DB (overlay). Each actor must pass through the Frontend server to retrieve information. When Front End Server receives a request, it queries the appropriate Shard, retrieve and process data, and replies with an aggregated result. It can also access web data via public APIs to get some useful information (like weather and roads traffic conditions).

### 3.2. Case study: Hospitals Supply Chain (HSC)

Our case study concerns hospitals supply chain, in which we have to deal with the following constraints:

– Stakeholders over the supply chain are geographically distributed, and each one has a specific need to some information in real-time, for example the customer (hospital) needs to get delivery dates and possible delivery ways of medications packages, food or covers; The carrier needs to know the position of its trucks and road traffic state from an external source like internet; The supplier would like to find the carrier that proposes the cheapest and fastest shipping service, etc.
– They should get necessary information without calling or emailing each other.
– Data flow is generated over the supply chain by an increasing number of stakeholders and customers, and potentially millions of sensors and tracked objects. The most part of this data is unstructured and then need a flexible schema data store.

According to these constraints, we integrated our DCLP with HSC, in order to automate and standardize data exchange between stakeholders, and break obstacles to improve cooperation.
The Logistics Information System architecture that we propose relies on MongoDB NoSQL database, as shown in Figure 6. The aim of this framework is to handle massive real-time and distributed data exchange and requests generated in a multi-actor logistics environment and provide logistics stakeholders like suppliers, carriers, customers or manufacturers with relevant information that they need at the right time.

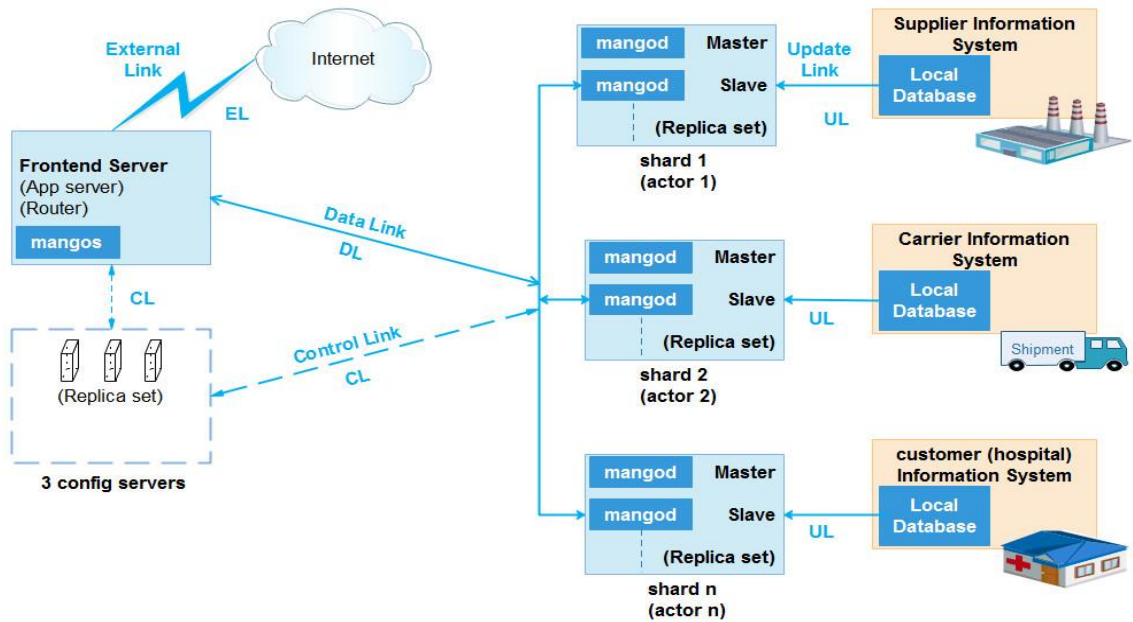

**Figure 6:** NoSQL Logistics Information System Architecture.

The first component of the framework is the *frontend server* or application server, wherein *MongoS* (MongoDB Shard) is installed. Mongos is a routing service for mongoDB shard configurations which aims to link between the application layer and the sharded cluster in order to determine the location of the data. When mongos receives a query, it communicates with the config server to get the information needed to forward the query to the right shard (replica set).

The second component is the *Config Server*, which serves to store the mapping or metadata that links requested data with the shard that contains it. In production environment, sharded clusters have exactly 3 config servers to ensure high availability and redundancy.

The third component is composed of shards which are used to store stakeholders' available data. A single shard is usually composed of a replica set with one Master and many slaves to ensure availability and redundancy. Each shard is connected to a local database of a logistic actor in order to collect necessary information via periodical updates.

Communication between the previous three main components is performed using data, control, update and also external links:

– Data link (DL) allows data forwarding and broadcasting in the global supply chain
– Control link (CL) allows transmission of requests and logistics network mapping between all framework components
– Update link (UL) enables feeding the shards of the distributed database
– External link (EL) collects information from the web on demand, which may be weather or traffic information, and so on

Data confidentiality aspect is also very important; the idea here is to provide management of the Frontend server to a trusted third party. Data made available voluntarily by stakeholders to feed shards constitute the minimum necessary to operate the system, it's the condition to enter the DCLP, we notice in this case a virtuous circle. No one has access to others' information directly; When an actor needs an information, he sends a request to the frontend server, this one will contact the config servers via Control Links to get the locationss of needed data, afterwards it get necessary data from all related shards or external sources; thus, it aggregates the data and uses the result to respond to the initial request via Data Link.

### 3.3. DCLP prototype

A prototype of our Distributed Cooperative Logistics Platform is being developed on IT resources of LITIS lab. It is fed by data from Supply Chains of Normandy hospitals, obtained through our partners in PERFAD project (Performance of decentralized supply chains). As we have static information (annual data of orders sorted by product families and sub-families, suppliers, stock locations), we developed from this information a data generator that can be used to feed the stakeholders' information systems with continuous real-time data flows, and these latter feed the associated MongoDB databases, and thus feed our prototype. Queries can be issued later by each actor, in order to verify the feasibility of such system, that aims to optimize logistics decisions made by actors, taking into account the freshness and obsolescence of certain data, such as expiry dates and traceability data for moving objects.

## 4. Conclusion and future scope of work

In this paper we proposed a NoSQL based Distributed Cooperative Logistics Platform (DCLP) framework, in order to improve cooperation between logistic stakeholders, and manage global supply chain data in real-time. Then, each actor can send requests to the frontend server and directly get related information without accessing to others' sensitive data. Cooperation and controlled information sharing between stakeholders is the solution to improve Supply Chains and decision making process of logistics stakeholders.

In future work, we will finish the development of the prototype, generalize it, and use DCLP as a decision making support tool. If feasibility is proven, our DCLP would guarantee the confidentiality of data through a trusted third party, facilitate the exchange of information between actors, and thus improve integration between Information Systems of these actors. Therefore, it would meet the good performance criteria of a Supply Chain.

### Acknowledgment

This work was financed by French government, region of Normandy, and European Union (European Regional Development Fund) within CLASSE and PERFAD projects.

## 5. References

1. Chen, D. Q., Preston, D. S. & Xia, W. Enhancing hospital supply chain performance: A relational view and empirical test. *J. Oper. Manag.* **31,** 391–408 (2013).
2. Sahay, B. S. & Ranjan, J. Real time business intelligence in supply chain analytics. *Inf. Manag. Comput. Secur.* **16,** 28–48 (2008).
3. Yu, X., Li, P. & Li, S. Research on data exchange between heterogeneous data in logistics information system. in *Communication Systems, Networks and Applications (ICCSNA), 2010 Second International Conference on* **1,** 127–130 (IEEE, 2010).
4. Duclos, L. K., Vokurka, R. J. & Lummus, R. R. A conceptual model of supply chain flexibility. *Ind. Manag. Data Syst.* **103,** 446–456 (2003).
5. Christopher, M. *Logistics and Supply Chain Management: Strategies for Reducing Cost and Improving Service Financial*. (Taylor & Francis, 1998).
6. Biswas, S. & Sen, J. A Proposed Framework of Next Generation Supply Chain Management Using Big Data Analytics. in *Proceedings of National Conference on Emerging Trends in Business and Management: Issues and Challenges* (2016).
7. Stephens, S. Supply Chain Operations Reference Model Version 5.0: A New Tool to Improve Supply Chain Efficiency and Achieve Best Practice. *Inf. Syst. Front.* **3,** 471–476 (2001).
8. Biswas, S. & Sen, J. *A Proposed Architecture for Big Data Driven Supply Chain Analytics*. (Social Science Research Network, 2016).
9. Hans Wortmann, Alblas, A. A., Buijs, P. & Peters, K. Supply Chain Integration for Sustainability Faces Sustaining ICT Problems. in *Advances in Production Management Systems. Sustainable Production and Service Supply Chains* (eds. Prabhu, V., Taisch, M. & Kiritsis, D.) 493–500 (Springer Berlin Heidelberg, 2013).
10. Manthou, V., Vlachopoulou, M. & Folinas, D. Virtual e-Chain (VeC) model for supply chain collaboration. *Int. J. Prod. Econ.* **87,** 241–250 (2004).

11. Kong, L. & Wu, J. Collaboration attitude choice-based intelligent production control model in dynamic supply chain system. in *IEEE International Conference Mechatronics and Automation, 2005* **3,** 1386–1390 Vol. 3 (2005).
12. Grolinger, K., Higashino, W. A., Tiwari, A. & Capretz, M. A. Data management in cloud environments: NoSQL and NewSQL data stores. *J. Cloud Comput. Adv. Syst. Appl.* **2,** 22 (2013).
13. Chang, F. *et al.* Bigtable: A distributed storage system for structured data. *ACM Trans. Comput. Syst. TOCS* **26,** 4 (2008).
14. DeCandia, G. *et al.* Dynamo: amazon's highly available key-value store. *ACM SIGOPS Oper. Syst. Rev.* **41,** 205–220 (2007).
15. Brewer, E. A. Towards robust distributed systems. in *PODC* **7,** (2000).
16. Gilbert, S. & Lynch, N. Brewer's Conjecture and the Feasibility of Consistent, Available, Partition-tolerant Web Services. *SIGACT News* **33,** 51–59 (2002).
17. Fox, A., Gribble, S. D., Chawathe, Y., Brewer, E. A. & Gauthier, P. Cluster-based Scalable Network Services. in *Proceedings of the Sixteenth ACM Symposium on Operating Systems Principles* 78–91 (ACM, 1997). doi:10.1145/268998.266662
18. Pritchett, D. BASE: An Acid Alternative. *Queue* **6,** 48–55 (2008).
19. Vogels, W. Eventually Consistent. *Commun ACM* **52,** 40–44 (2009).
20. He, C. Survey on NoSQL Database Technology. *J. Appl. Sci. Eng. Innov. Vol* **2,** (2015).
21. Nyati, S. S., Pawar, S. & Ingle, R. Performance evaluation of unstructured NoSQL data over distributed framework. in *2013 International Conference on Advances in Computing, Communications and Informatics (ICACCI)* 1623–1627 (2013). doi:10.1109/ICACCI.2013.6637424
22. Zahid, A., Masood, R. & Shibli, M. A. Security of sharded NoSQL databases: A comparative analysis. in *2014 Conference on Information Assurance and Cyber Security (CIACS)* 1–8 (2014). doi:10.1109/CIACS.2014.6861323
23. Haughian, G., Osman, R. & Knottenbelt, W. J. Benchmarking Replication in Cassandra and MongoDB NoSQL Datastores. in *Database and Expert Systems Applications* (eds. Hartmann, S. & Ma, H.) 152–166 (Springer International Publishing, 2016).